\documentstyle[12pt]{article}

\begin{document}

\begin{center}

{\Large {\bf Auxiliary potential in no-core shell-model calculations}}

\vspace{0.2in}

D. C. Zheng$^1$, B. R. Barrett$^1$, J. P. Vary$^2$, and H. M\"uther$^3$\\

\end{center}
\begin{small}

\noindent
$^1${\it Department of Physics, University of Arizona,
	Tucson, Arizona 85721}

\noindent
$^2${\it Department of Physics and Astronomy, Iowa State University,
	Ames, Iowa 50011}

\noindent
$^3${\it Institut f\"ur Theoretische Physik, Universit\"at T\"ubingen,
D-72076 T\"ubingen, Germany}
\end{small}

\vspace{0.5in}

\begin{abstract}
The Lee-Suzuki iteration method is used to include the
folded diagrams in the calculation of the
two-body effective interaction $v^{(2)}_{\rm eff}$
between two nucleons in a no-core model space.
This effective interaction still depends upon
the choice of single-particle basis utilized in the shell-model
calculation. Using a harmonic-oscillator single-particle basis and the
Reid-soft-core {\it NN} potential, we find that $v^{(2)}_{\rm eff}$
overbinds $^4\mbox{He}$ in 0, 2, and $4\hbar\Omega$ model spaces.
As the size of the model space increases, the amount of overbinding
decreases significantly. This problem of overbinding in small model
spaces is due to neglecting effective three- and
four-body forces. Contributions of
effective many-body forces are suppressed by using the Brueckner-Hartree-Fock
single-particle Hamiltonian.
\end{abstract}

\pagebreak

\section{Introduction}
Previous calculations of the shell-model effective interaction involved
a number of uncertainties. The major ones included
the choice of the single-particle (s.p.) basis, the choice of the starting
energy $\omega$ in the Brueckner $G$-matrix \cite{G}
calculation, the neglected
contribution from higher-order core-polarization diagrams, and
the effects of both the real and effective
three- and higher-body forces.
It was proposed in Ref.\cite{nc} that the
 core-polarization diagrams could be eliminated by adopting
a no-core model space, in which all the nucleons in a nucleus are treated
as active. In Refs.\cite{lighta,mfd}, the no-core approach has been
used and satisfactory results are obtained for light nuclei.

The results of this no-core approach depend on the choice of the starting
energy for the $G$-matrix. One may argue that this
uncertainty in the choice of the starting energy $\omega$ can
be avoided by evaluating an energy-independent interaction employing the
iteration methods proposed by Lee and
Suzuki \cite{ls} and Krenciglowa and Kuo \cite{kuo}. Applying this
technique, one can sum the folded diagrams to all orders to
obtain a starting-energy-independent effective interaction
$v_{\rm eff}^{(2)}$.
Therefore, with the use of a no-core space and including the folded
diagrams, the effective interaction obtained is subject to only two
of the major uncertainties mentioned above, i.e.,
the choice of the s.p.~basis and the effect of
the neglected effective three- and higher-body forces.

These two remaining uncertainties are related.
With an optimally chosen s.p.~basis, the contribution from the
effective many-body forces could be minimized \cite{barrett}.
Furthermore, both of the uncertainties are related to the size of
the model space and are expected to diminish as the size
increases. Indeed, the effective interaction is only introduced with the
truncation of the infinite Hilbert space to a finite-size model space.
Furthermore, one should keep in mind that
so-called Q-box diagrams, like the one displayed in
Fig.~1a, are not the only source of effective many-body forces.
Even if the Q-box is restricted to two-body terms, which means
the $G$-matrix for no-core calculations, the inclusion of folded
diagrams yields effective three-body forces (see Fig.~1b). It has been
demonstrated that such many-body forces are non-negligible \cite{3bf},
in particular if a large number of active particles has to be
considered \cite{3bfp}. As we will discuss below,
one may try to minimize the effects of such many-body forces by
introducing an appropriate auxiliary field. One may view the present
effort as an extension of the work of Ref.\cite{barrett}
to the case of realistic $NN$ potentials which capitalizes on the
results presented in Refs.\cite{3bf,3bfp}.

In this work, we will use the Lee-Suzuki method \cite{ls}
to calculate the starting-energy-independent two-body
effective interactions $v_{\rm eff}^{(2)}$ for
no-core, harmonic-oscillator (HO) model spaces and study
the dependence of the shell-model results obtained for $^4\mbox{He}$
with $v_{\rm eff}^{(2)}$ on the HO basis parameter $\hbar\Omega$ and
the size of the model space.
It has been noticed in a previous work \cite{veff} that
$v^{(2)}_{\rm eff}$ tends to overbind light nuclei.
Here we will show that the overbinding is quite significant when the
model space is relatively small. We will show that the overbinding
problem can be cured by introducing an auxiliary field, such as the
Brueckner-Hartree-Fock (BHF) approximation.

After this introduction we will present some details on the evaluation
of the energy-independent effective two-body force for no-core
shell-model calculations in section 2. Numerical results for the
binding energy of $^4\mbox{He}$ will be presented in section 3. In
section 4 we will discuss the influence of an auxiliary potential, and
section 5 contains the conclusions of the present investigation.

\section{Calculation of $v_{\rm eff}^{(2)}$}
The Brueckner $G$ matrix is calculated according to the following equation:
\begin{equation}
G(\omega) = v_{12} + v_{12} \frac{Q}{\omega - (h_1+h_2+v_{12})} v_{12},
					\label{g}
\end{equation}
where $v_{12}$ is the $NN$ force for which we will use the
Reid-soft-core (RSC) potential \cite{rsc}, $\omega$ is the starting energy,
$Q$ is the Pauli operator which excludes the scattering into the
two-particle states inside the model space. For a full no-core $N\hbar\Omega$
space, we define $Q$ as
\begin{eqnarray}
Q &=& 0 \hspace{0.2in} {\rm for} \hspace{0.1in} n_1+n_2 \leq N \nonumber \\
  &=& 1 \hspace{0.2in} {\rm for} \hspace{0.1in} n_1+n_2 > N .
			\label{Pauli}
\end{eqnarray}
where $n_i=2n_r(i)+l(i)=0, 1, \ldots,$
are the principal quantum numbers of the s.p.~states
occupied by the two intermediate-state nucleons in the multiple
scattering process.

For the first part of our discussion (see section 4 for an alternative
choice) the s.p.~Hamiltonian in Eq.(\ref{g}) is taken as
\begin{equation}
h_i= t_i + u_i =
 \frac{\mbox{\boldmath $p$}_i^2}{2m} + (\frac{1}{2}m\Omega^2r_i^2 - V_0)
   =  t_i + (u_i^{\rm HO}-V_0) = h_i^{\rm HO}-V_0,
\end{equation}
where we use a s.p.~potential ($u_i$) that is of the shape
of a harmonic oscillator but is shifted downward by an amount $V_0$
to make it more realistic.
The quantity $V_0$ represents the depth of the mean field of the nuclear
medium. It is convenient to define a shifted starting energy as
$\omega' = (\omega+2V_0)$ and rewrite Eq.(\ref{g}) as
\begin{equation}
G(\omega') = v_{12} + v_{12} \frac{Q}{\omega' -
	(h^{\rm HO}_1+h^{\rm HO}_2+v_{12})} v_{12},
					\label{g'}
\end{equation}
where $h^{\rm HO} = t_i + u_i^{\rm HO}$ is now a pure HO Hamiltonian.
Since we will use the Lee-Suzuki iteration method \cite{ls} to
take into account the folded diagrams, the resulting effective interaction
will be independent of the starting energy $\omega'$ as well as the shift
$V_0$. Therefore, no specific choice for the value of $V_0$ needs to be made.

However, in the case when the folded diagrams were ignored (which is
a common practice in effective-interaction calculations), one would have
to choose a reasonable starting energy to minimize the
contribution from the folded diagrams. It should then be noted
that the starting energy $\omega'$ used in Eq.(\ref{g'}), unlike
$\omega$, does not correspond to the energy $E_2$ of the initial
two-particle state in the ladder diagrams. Rather, it is related to
$(E_2+2V_0)$. When the two nucleons in the initial state occupy
bound s.p.~states, $E_2$ is negative. But $\omega'\simeq (E_2+2V_0)$
could very well be positive. In fact, it has been found \cite{bhm,vy}
that for the two valence neutrons in $^{18}\mbox{O}$, a value of
about 70 MeV for $\omega'$ yielded reasonable $G$-matrix elements.

In order to obtain the starting-energy-independent two-body effective
interaction $v^{(2)}_{\rm eff}$, we calculate $G(\omega')$ of
Eq.(\ref{g'}) for 11 values of
$\omega'$ ranging from about $-5\hbar\Omega$ to about $5\hbar\Omega$.
These 11 sets of $G$ matrices are then used to numerically
calculate the derivatives of $G(\omega')$ with respect to $\omega'$
to the 9th order.
Once the derivatives of $G(\omega')$ are obtained,
we proceed with the Lee-Suzuki method to obtain $v^{(2)}_{\rm eff}$.
Here we point out that the number of iterations needed for convergence
strongly depends on the value of $\omega'$ at which the derivatives are
evaluated. It generally exceeds the number of derivatives retained in the
iteration procedure.

In Fig.2, we show the values of the diagonal two-body matrix elements (TBMEs)
of $G(\omega')$ and $v_{\rm eff}^{(2)}$ in the states
$|(0s_{1/2})^2\rangle_{J,T}$ with $J$=0, $T$=1 and $J$=1, $T$=0
for a wide range of $\omega'$. For these matrix elements we use
$\hbar\Omega$=16 MeV and the Pauli operator is defined in Eq.(\ref{Pauli})
with $N$=4. It can be seen from the figure that the matrix elements
of $G(\omega')$ decrease (i.e., become more attractive) with increasing
$\omega'$, while those of $v_{\rm eff}^{(2)}$ are independent of $\omega'$.

We also note in Fig.2 that there is a particular value of $\omega'$ for
which the matrix elements of $G(\omega')$ are about equal to those of $v_{\rm
eff}^{(2)}$.  This observation is the basis of an approximation scheme
presented and tested in Ref.\cite{veff} and then used in Ref.\cite{mfd}.

\section{Shell-Model Results}
We perform the matrix diagonalization for the
 shell-model Hamiltonian
\begin{equation}
H_{\rm SM} = \left(\sum_{i=1}^A t_i -T_{\rm c.m.}\right)
	+ \sum_{i<j}^A v^{(2)}_{\rm eff}(ij) + V_{\rm Coulomb}
	+ \lambda (H_{\rm c.m.}-\frac{3}{2}\hbar\Omega) .	\label{hsm}
\end{equation}
In the above equation the $t_i={\bf p}_i^2/(2m)$
are the one-body kinetic energies,
$T_{\rm c.m.}=(\sum_i {\bf p}_i)^2/(2mA)$ is
the c.m. kinetic energy and $V_{\rm Coulomb}$ is the Coulomb
interaction. The proton and neutron masses are taken to be the same.
The last term (with $\lambda$=10) in the above equation forces
the c.m. motion of the low-lying states in the calculated spectrum
to be in its lowest HO configuration.

In Fig.3, we plot the calculated ground-state (g.s.) energy of $^4\mbox{He}$
as a function of the HO basis parameter $\hbar\Omega$ for three
model spaces of different sizes, $0\hbar\Omega$ (``$N$=0'' curve),
$2\hbar\Omega$ (``$N$=2'' curve), and $4\hbar\Omega$ (``$N$=4'' curve).
In the $0\hbar\Omega$ model space which consists
of only the $0s_{1/2}$ major shell,
the g.s.~energy begins at -33.9 MeV for $\hbar\Omega$=10 MeV,
decreases to a minimum of -43.3 MeV at $\hbar\Omega$=22 MeV
and then increases to -41.6 MeV for
$\hbar\Omega$=28 MeV. These results significantly overbind the g.s.~of
$^4\mbox{He}$. They are 5.6 to 15.0 MeV lower than
the experimental g.s.~energy of -28.3 MeV and
are 9.3 to 18.7 MeV lower than the value of -24.6 MeV
obtained in the (nearly exact) Green's function Monte Carlo (GFMC)
approach \cite{carlson} using the RSC potential.

In this one-major-shell model space,
there is a simple way to obtain the above results.
The g.s.~energy of $^4\mbox{He}$ can be expressed in
terms of the effective-interaction TBMEs as ($0s\equiv 0s_{1/2}$)
\begin{equation}
E_{\rm gs} = 3\left(\frac{3}{4}\hbar\Omega \right)
+3\left[\langle 0s^2|v^{(2)}_{\rm eff}|0s^2\rangle_{J=0,T=1}
+\langle 0s^2|v^{(2)}_{\rm eff}| 0s^2\rangle_{J=1,T=0}\right],
\label{eq:egs}
\end{equation}
where the first term is the kinetic energy (with the c.m.~contribution
subtracted) and the second term is the effective-interaction energy.
According to Ref.\cite{zvb}, we know that for this one-dimensional
model space, the effective-interaction TBMEs
are related to the eigenenergies of the Schr\"odinger equation:
\begin{equation}
(h_1+h_2 + v_{12}) \phi_{J,T} = E_{J,T}\, \phi_{J,T}      \label{schr}
\end{equation}
through
\begin{equation}
\langle 0s^2 |v^{(2)}_{\rm eff}| 0s^2\rangle_{J,T}
 = E_{J,T} - 2\left( \frac{3}{2}\hbar\Omega\right),       \label{veff}
\end{equation}
where $\frac{3}{2}\hbar\Omega$ is the eigenenergy of the
s.p.~Hamiltonians $h_1$ and $h_2$ for the $0s_{1/2}$ state.
For $\hbar\Omega$=22 MeV, we obtained
$$E_{0,1} = 55.183 \, {\rm MeV} \hspace{0.2in} {\rm and} \hspace{0.2in}
  E_{1,0} = 45.892 \, {\rm MeV},$$
which lead to
$$\langle 0s^2|v^{(2)}_{\rm eff}|0s^2\rangle_{0,1}
= -10.817 \, {\rm MeV}$$
and
$$\langle 0s^2 |v^{(2)}_{\rm eff}|0s^2\rangle_{1,0}
= -20.108 \, {\rm MeV}.$$
Therefore,
$$E_{\rm gs} = 3\left(\frac{3}{4}\times 22 \right)+3(-10.817-20.108)
= -43.275 \, {\rm MeV},$$
which agrees with the result that we obtained through the Lee-Suzuki
iteration procedure.
We further remark that the effective-interaction TBMEs
$\langle 0s^2|v^{(2)}_{\rm eff}|0s^2\rangle_{J,T}$
are equal to $G_{J,T}(\omega')$ with $\omega'=E_{J,T}$. This is
another property observed in Ref.\cite{zvb} for the effective interaction
in a one-dimensional model space. From these arguments we can see that
the overbinding in this very limited model space is related to
the fact that the matrix elements $G_{J,T}(\omega'=E_{JT})$ are very attractive
for the rather positive values of $E_{0,1}$ and $E_{1,0}$.

Note that in the limit of $\hbar\Omega=0$, the one-dimensional model-space
result for the g.s.~energy of $^4\mbox{He}$ is
\begin{equation}
E_{\rm gs} = 3(0-2.2246) = -6.6738 \, {\rm MeV}   \hspace{0.2in}
				({\rm for}\; \; \hbar\Omega=0),
\end{equation}
where -2.2246 MeV and 0 are the lowest eigenenergies of the two-body system
with $J$=1, $T$=0 (deuteron) and $J$=0, $T$=1, respectively.
This (-6.6738 MeV) is the limit that the ``$N$=0'' curve in Fig.2
will approach as $\hbar\Omega \rightarrow 0$.

In the $2\hbar\Omega$ model space, the results for the binding energy of
$^4\mbox{He}$ are reduced considerably (see Fig.3) although
they are still larger than the experimental value as well as
the more exact theoretical value for the RSC potential.
Note that the value of $\hbar\Omega$ at which the lowest g.s.~energy is
obtained
is between 16 MeV and 18 MeV in the $2\hbar\Omega$ model space
($N$=2). This is quite different from corresponding value of $\hbar\Omega$=22
MeV with the $0\hbar\Omega$ space ($N$=0). At $\hbar\Omega$=16 MeV, the
$2\hbar\Omega$ result for the binding energy is 33.6 MeV, which still overbinds
the g.s.~by a large amount. The reduction of the calculated energy with
increasing model space can easily be understood from the following
observations:
If the model space is increased, the Pauli operator $Q$ in the Bethe-Goldstone
Eq.(\ref{g}) ensures that the energy $\omega_{1}$ of the lowest pole in $G$ is
shifted to higher energies, as this energy correspond to the energy of the
lowest 2 particle state outside the model space. Therefore the matrix elements
of $G$ calculated at the same starting energy are less attractive for
the larger model space.
The net effect of enlarging the model space with the appropriately recalculated
effective interaction at the two-particle level is to reduce the overbinding.

The results continue to improve as we increase the model space
from $2\hbar\Omega$ to $4\hbar\Omega$. Now the lowest g.s.~energy
of about -31.9 MeV is found when $\hbar\Omega$ falls between 14 and 16 MeV.
This lowest energy is about 1.7 and 11.4 MeV higher than the corresponding
values of -33.6 MeV and -43.3 MeV for the $2\hbar\Omega$ and
$0\hbar\Omega$ spaces, respectively.

The dependence of the results on the $\hbar\Omega$ value also
weakens substantially as we go from $2\hbar\Omega$ to
$2\hbar\Omega$ and to $4\hbar\Omega$.
We can quantify this dependence by defining a dimensionless parameter
\begin{equation}
C_a \equiv {\rm Average}\hspace{0.1in}{\rm of}\hspace{0.15in}
	\frac{|E_{\rm gs}(\hbar\Omega+2)-E_{\rm gs}(\Omega)|}{2 {\rm MeV}}
\end{equation}
that characterizes the ``average curvature''.
For the $N$=0 curve in Fig.3, $C_a$ is 0.61. It
decreases to 0.39 for the $N$=2 curve; and it further reduces to
0.27 for the $N$=4 curve.
Ultimately, when an infinite Hilbert space is used,
the results for the g.s.~energy should show a complete
independence of $\hbar\Omega$ (i.e., $C_a$ defined above
vanishes) and should converge to the exact result of $-24.55$ MeV
\cite{carlson}. This is indeed the trend we are seeing in the
$0\hbar\Omega$, $2\hbar\Omega$ and $4\hbar\Omega$
calculations but there is still a considerable gap between
the $4\hbar\Omega$ results and the converged, exact value.
Crude extrapolation of the $C_a$ values for $N$=0, 2 and 4
indicates that one may have to do an $8\hbar\Omega$ calculation
in order to reduce $C_a$ to less than 0.1, at which point
a 10 MeV change in $\hbar\Omega$ will, on average, results in less than
1 MeV change in the ground-state energy.

\section{Auxiliary single-particle potential}

The discrepancy between the energies obtained in the shell-model
calculations of the preceding section and the exact result obtained
for the RSC potential is due to the fact that some effective three- and
four-body  forces are ignored in our calculations.

Since shell-model calculations with inclusion of many-body forces are
rather involved \cite{3bfp}, we would prefer to find a way to diminish
the effect of these many-body terms. For that purpose we
consider the lowest-order contribution to the three-body folded
diagrams displayed in Fig.1b. The contribution of this effective
three-body force to the binding energy within the $0\hbar\Omega$ model
space is represented by the diagram displayed in Fig.1c. If we
introduce an auxiliary s.p.~potential for all active states
of the model space according to the BHF choice
\begin{equation}
\epsilon_{i}^{\rm BHF} = t_{i} + \sum_{J,T} \frac{(2J+1)(2T+1)}
{2(2j_{i}+1)} \langle (i\, 0s)|G^T(\omega = \epsilon_{i} + \epsilon_{0s})|
(i\, 0s)\rangle_{J,T} \; , \label{eq:bhf}
\end{equation}
the contribution of the three-body term of Fig.1c and many higher-order
diagrams originating from folding would be canceled by corresponding
diagrams with s.p.~insertions. Therefore, in this section,
we would like to discuss the influence of introducing this auxiliary
potential on the binding energy calculated in a $0\hbar\Omega$ model
space.

It should be noted that Eq.(\ref{eq:bhf}) defines an auxiliary
potential for the states within the model space. For s.p.~states
outside the model space we assume pure kinetic energy. This
means that the Bethe-Goldstone Eq.(\ref{g}) has been solved with the
kinetic energy ($h_{i}=t_{i}$) for the s.p.~spectrum of the
s.p.~states outside the model space. The resulting $G$ matrix is
denoted by $G^T$, as we already did in Eq.(\ref{eq:bhf}).
When one comes to calculate the BHF s.p.~energies as defined in
Eq.(\ref{eq:bhf}), this new choice for the intermediate energy spectrum
is preferred, since our previous choice
involves an unspecified shift $V_0$ in the one-body potential, which makes
it difficult to unambiguously relate the starting energy with
the BHF s.p.~energies in Eq.(\ref{eq:bhf}).

In order to appreciate the significant effects of
using the auxiliary potential, which we will soon discuss, we need to
separate the influence of the new choice for
the intermediate spectrum on the results. To this end,
we first evaluate the binding energy of $^4$He in the
$0\hbar\Omega$ model space without assuming an auxiliary potential.
This means that we first solve Eq.(\ref{schr}) with $h_{i}=t_{i}$ for
the new G-matrix
\begin{equation}
\left[t_1+t_2 + G^T(\omega =E_{J,T})\right] \phi_{J,T} = E_{J,T}\,
\phi_{J,T} \; ,     \label{schrp}
\end{equation}
so as to determine the matrix elements
\begin{equation}
\langle 0s^2 |v^{(2)}_{\rm eff}|0s^2\rangle_{J,T} =
\langle 0s^2 |G^T(\omega =E_{J,T})|0s^2\rangle_{J,T} \label{eq:vef}\; .
\end{equation}
We then evaluate the energy according to Eq.(\ref{eq:egs}) and add the
Coulomb repulsion between the two protons. Results for
this calculation without an auxiliary potential are displayed in Fig.4
(solid line -- ``Without aux. pot.'').
These results are essentially the same as those we
obtained in the previous section (Fig.3), indicating
that the results are rather insensitive to the choice
of the intermediate energy spectrum in the Bethe-Goldstone equation.
This is also consistent with an earlier study \cite{jaqua} in
which the role of the single-particle potential was examined in some detail at
the level of two-body effective-interaction calculations.  Here we extend
those results to the case of starting-energy-independent effective two-body
interactions.  This is important since, by eliminating the issue of whether the
single-particle insertions on intermediate particle lines are responsible for
overbinding, we are then forced to consider the effective many-body forces that
are addressed in the present study.

We now consider the BHF choice for the s.p.~potential. We again
use the one-dimensional model space for simplicity.
In analogy to Eq.(\ref{veff}), we now have
\begin{equation}
\left[ 2 \epsilon_{0s}^{\rm BHF} + \langle 0s^2 |G^T(\omega =E_{J,T})
|0s^2\rangle_{J,T} \right] \phi_{J,T} = E_{J,T}\,
\phi_{J,T}\; ,      \label{schrq}
\end{equation}
which determines the matrix elements of $v^{(2)}_{\rm eff}$ according to
Eq.(\ref{eq:vef}). Results for the binding energy of $^4$He with this
choice of the auxiliary potential are also displayed in Fig.4 by the
solid line labeled ``With BHF pot.''. One observes that the use of
the BHF auxiliary potential, which has been introduced to measure the
effects of terms like the three-body folded diagrams of Fig.1b, reduces
the calculated binding energy drastically. Even the minimal value for
the binding energy ($E_{\rm gs}$=-17.9 MeV), which is obtained around
$\hbar\Omega$ = 16 MeV, is well above the ``exact result'' of
Ref.\cite{carlson}.

For the sake of comparison, we also show in Fig.4 the results (dotted line)
for the BHF approximation (restricting the s.p.~wave functions to
the HO wave functions for a given $\hbar\Omega$),
which can be obtained by replacing
$E_{J,T}\rightarrow 2 \epsilon_{0s}^{\rm BHF}$. This approximation
would correspond to a $0\hbar\omega$ no-core calculation, assuming the
BHF auxiliary potential but ignoring the effects of two-body folded
diagrams. We see that the two-body folded diagrams yield
a repulsion of about 2 MeV. Such a repulsive effect has also
been observed in shell-model calculations within the $sd$ shell \cite{3bf}.

\section{Conclusions}
Energy-independent effective two-body interactions
$v^{(2)}_{\rm eff}$ are determined
to calculate the g.s.~energy of $^4\mbox{He}$
in three no-core HO model spaces with $\hbar\Omega$
ranging from 10 to 28 MeV. The results overbind the g.s.~of $^4\mbox{He}$
for all the three spaces and  for all the $\hbar\Omega$ values that we
have used. The amount of overbinding is largest in the
$0\hbar\Omega$ space and decreases as we increase the size of the model space
from $0\hbar\Omega$ to $2\hbar\Omega$ and to $4\hbar\Omega$.
The dependence of the calculated g.s.~energy on the value of
$\hbar\Omega$ also becomes weaker as the size of the model space
increases. However, even in the $4\hbar\Omega$ calculation, the
lowest g.s.~energy of -31.9 MeV, which
we obtained with $v^{(2)}_{\rm eff}$ at
$\hbar\Omega$=16 MeV, still overbinds $^4\mbox{He}$ by
3.6 MeV when compared with the experimental binding energy and
by 7.3 MeV when compared with the binding energy obtained in the
GFMC approach \cite{carlson}.

This overbinding is caused by the fact that we are neglecting some
effective many-body forces. Our results using a BHF auxiliary
potential, which includes certain effective many-body-force terms,
show significant effects on no-core shell-model calculations,
especially when the model space is small.
More studies are required to find an optimal auxiliary potential,
which minimizes the effects of such many-body forces.

\section*{Acknowledgment}
Two of us (B.R.B. and D.C.Z.) acknowledge
partial support of this work by the National Science Foundation,
Grant No.~PHY-9321668. One of us
(J.P.V.) acknowledges partial support by the U.S.~Department of
Energy under Grant No.~DE-FG02-87ER-40371, Division
of High Energy and Nuclear Physics.

\vspace{0.2in}

\begin{small}

\end{small}

\pagebreak

\section*{Figure captions}

\noindent
{\bf Fig.1} \hspace{0.1in}
Contributions to the effective three-body force originating from the
so-called Q-box (a) and folded diagrams (b). The diagram (c) represents
the contribution of diagram(b) to the ground-state energy within a
$0\hbar\Omega$ model space. The bare lines refer to s.p.~states
within the model space while the ``railed'' lines respresent
s.p.~states outside the model space.

\vspace{0.2in}

\noindent
{\bf Fig.2} \hspace{0.1in}
The two-body matrix elements, $\langle 0s^2 | G(\omega')|0s^2\rangle_{J,T}$
(solid lines) and $\langle 0s^2 | v_{\rm eff}^{(2)}|0s^2\rangle_{J,T}$
(dotted lines) for the $4\hbar\Omega$ model space with $\hbar\Omega$=16 MeV
as a function of the starting energy $\omega'$.
Note that the TBMEs of $v_{\rm eff}^{(2)}$ are independent of $\omega'$.

\vspace{0.2in}

\noindent
{\bf Fig.3} \hspace{0.1in}
The ground-state energy of $^4\mbox{He}$ obtained from
$v_{\rm eff}^{(2)}$ as a function of $\hbar\Omega$
and $N$, the size of the model space.

\vspace{0.2in}

\noindent
{\bf Fig.4} \hspace{0.1in}
The g.s.~energy of $^4\mbox{He}$ as a function of $\hbar\Omega$
obtained from the $0\hbar\Omega$ calculation. Results
are represented for using kinetic energies
(labeled: Without aux. pot.) or BHF s.p.~energies (label:
With BHF pot.) as an auxiliary potential. For a comparison the
dashed line displays the energies obtained with the BHF approximation.

\end{document}